\documentclass[aps,prl,twocolumn,superscriptaddress]{revtex4}
\usepackage{amsmath}
\usepackage{amssymb}
\usepackage{graphicx}

\begin{document}

\title{Tunneling into strongly biased Tomonaga-Luttinger liquid}

\author{Maxim Trushin} 
\affiliation{1. Institut f\"ur Theoretische Physik, Universit\"at Hamburg,
Jungiusstr 9, D-20355 Hamburg, Germany}
\affiliation{1. Institut f\"ur Theoretische Physik, Universit\"at
Regensburg, D-93040 Regensburg, Germany}
\author{A. L. Chudnovskiy}
\affiliation{1. Institut f\"ur Theoretische Physik, Universit\"at Hamburg,
Jungiusstr 9, D-20355 Hamburg, Germany}

\date{\today}

\begin{abstract}
We calculate the tunneling density of states for a Tomonaga-Luttinger liquid placed under a strong bias voltage. For the tunneling  through a side-coupled point contact, one can observe the power law singularities in the tunneling density of states separately for the right- and left-movers despite the point-like tunnel contact. Deviations of the nonequilibrium tunneling exponents from the equilibrium case are discussed.
\end{abstract}

\pacs{71.10.Pm}

\maketitle

{\em Introduction.}
Power-law suppression of the tunneling density of states into the Tomonaga-Luttinger liquid is one of the most profound manifestations of interactions in one dimensional (1D) electron systems  \cite{precursors}. It paves the way to get the information about interactions in 1D systems experimentally \cite{experiment-carbon,experiment-semi}.
Exact calculation of the tunneling density of states is possible
in the framework of the Tomonaga-Luttinger liquid (TLL) model using the bosonization 
method \cite{classics,vonDelft}. 
The basis of the TLL model is the linearization of one particle spectrum around the 
right and left Fermi points. 
At the same time, a proper account for the finite curvature of the one particle dispersion is necessary for theoretical description of high energy excitations in 1D electron systems and in particular for the description of strongly nonequilibrium 1D systems 
\cite{curvature}.

\begin{figure}
\includegraphics[width=\columnwidth]{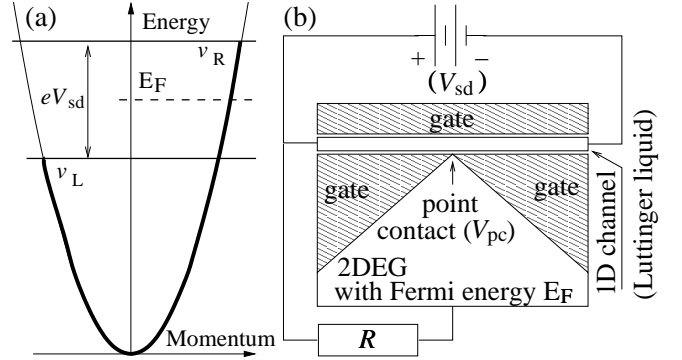}
\caption{\label{fig1}
(a) Occupation of the dispersion parabola in 1D channel
in presence of the source--drain voltage $V_\mathrm{sd}$ at 
negligible $V_\mathrm{pc}$.  The Fermi energy of the 2DEG lies nearly in
the middle between quasi Fermi levels of the biased TLL.
(b) Schematics of the device proposed.}
\end{figure}

In this letter we consider tunneling from the Fermi-liquid reservoir into the nonequilibrium TLL through a point tunnel contact (see Fig.~\ref{fig1}).  The nonequilibrium conditions are created by a strong transport voltage $V_\mathrm{sd}$ applied to a TLL channel.
In the equilibrium the whole system is filled by electrons up to the Fermi energy $E_F$.
Finite source-drain voltage  results in
the shift of the chemical potentials for the right- and left-movers to the quasi-Fermi energies  $E_F+eV_\mathrm{sd}/2$ and $E_F-eV_\mathrm{sd}/2$ respectively.  At strong enough voltages, the nonlinearity of the electronic dispersion leads to different Fermi velocities of the right- and left-movers $v_{R,L}=\sqrt{(2E_F\pm eV_\mathrm{sd})/m^*}$, as depicted in Fig.~\ref{fig1}. (Here, $m^*$ is the effective electron mass.) In turn, the tunneling densities of states for the left- and right-moving spectral branches differ.   Furthermore, since the direction of partial tunneling currents into the left branch and out of the right branch are opposite, these two tunnel currents do not compensate any more even at zero voltage $V_\mathrm{pc}$ at the point contact (see Fig.~\ref{fig1}). Therefore, a finite tunnel current flows between the nonequilibrium TLL and the reservoir. This current depends as a power law both on the source drain voltage in TLL $V_{\mathrm {sd}}$ and on the voltage on the point contact $V_{\rm pc}$, with the exponent reflecting the interaction strength in TLL. The exponent differs  from the one describing the tunneling anomaly in the equilibrium TLL. We propose a modification of the bosonization approach that allows to calculate the nonequilibrium tunneling density of states in TLL analytically.  The dependence of the current through the point contact  on the source-drain and point contact voltages $I_{\rm pc}(V_{\rm sd}, V_{\rm pc})$ taking into account electron-electron interactions in the 1D channel is the main result of this paper. 

To provide a theoretical description of TLL under a strong source-drain voltage, we have to deal with quite unusual 1D electron liquid with different densities of states for opposite chiralities. To our knowledge, this modification of Tomonaga-Luttinger model has not been solved so far. The Hamiltonian for the TLL channel consists of a sum of kinetic and interaction energies 
\begin{equation}
\label{allterms}
H_{TLL}=H_{\mathrm{kin}}+H_{\mathrm{int}},
\end{equation}
where the kinetic energy term takes into account different quasi-Fermi energies and different 
Fermi velocities for the left- and right-movers 
\begin{eqnarray}
\nonumber &&
H_\mathrm{kin}=\int\frac{dx}{2\pi}
\left[\psi_L^\dagger(x)\left(-eV_\mathrm{sd}/2 + i\,\hbar v_L\frac{\partial }{\partial
x}\right)\psi_L(x)+\right.  \\
&&
\left.\psi_R^\dagger(x)\left(eV_\mathrm{sd}/2 -i\,\hbar v_R\frac{\partial}{\partial x}\right)
\psi_R(x)\right],
\label{Hkin}
\end{eqnarray}
and $H_{\rm int}$ is the standard charge density interaction Hamiltonian of TLL.
In Eq. (\ref{Hkin}) the energy is counted from the position of the Fermi energy in the unbiased TLL. For description of the tunneling it proves convenient to perform a gauge transformation $\psi_{R(L)}(x)\to \hat{U}(x)^\dagger
\psi_{R(L)}(x)\hat{U}(x)$. The matrix $\hat{U}$ can be represented as $\hat{U}(x)=\hat{U}_R(x) \hat{U}_L(x)$ with 
\begin{equation}
\hat{U}_{R(L)}(x)=\exp\left[-\frac{ix eV_\mathrm{sd}}{2\hbar v_{R(L)}}
\int dx' \psi^\dagger_{R(L)}(x')\psi_{R(L)}(x')\right].
\label{gauge}
\end{equation}
This gauge transformation absorbs the quasi Fermi energies into the spatial dependence of the transformed fields.

Using the bosonization identity,
the total Hamiltonian (\ref{allterms})  can be written in a bosonized 
form often encountered in the literature, namely
\begin{equation}
\label{startham}
H=\frac{\hbar v_R}{2} \int\frac{dx}{2\pi}
\partial_x {\phi}^T \left(\begin{array}{cc}
1 + g_4 & g_2 \\ 
g_2 &  g + g_4
\end{array} \right) \partial_x {\bf\phi},
\end{equation}
where ${\phi}^T=\left(\phi_R(x), \phi_L(x)\right)$  is the bosonic field corresponding to the fermions in TLL.
Here the terms with quadratic number operators  $N^2_{L,R}$ are omitted. 
The interaction constants can be found via Fourier transforms of a given interelectron potential \cite{gogolin}.
For the spinless case considered here we have $g_4=V(q=0)/\hbar v_R$ and 
$g_2=\left[V(q=0)-V(q=k_F^R-k_F^L)\right]/\hbar v_R$, where
$V(q)$ is the Fourier transform of the screened interelectron interaction potential,
and $k_F^{R,L}$ are the Fermi wave vectors.
The constant $g=v_L/v_R$ is assumed to be a measure of the dispersion law asymmetry.

{\em Diagonalization of the Hamiltonian.}
The conventional way to diagonalize the Hamiltonian of any interacting  system
in the framework of the Tomonaga-Luttinger model is to introduce
so-called ``dual fields''. The standard couple of ``dual fields'' is defined
as a difference and a sum between boson fields with opposite chiralities
\cite{vonDelft}.
However, because of the different Fermi velocities for right- and left-moving electrons, the 
introduction of dual fields  cannot be made in that direct manner for the problem at hand.  Rather, we introduce additional fictitious bosonic fields $\phi'_R(x)$ and $\phi'_L(x)$ in a way which
does not change the dynamics of the system.
Then, in terms of the four-component bosonic fields ${\mathcal B}=(\phi_R, \phi_L, 
 \phi'_L, \phi'_R)^T$,  the Hamiltonian (\ref{startham}) can be
rewritten as 
\begin{equation}
\label{main}
H=
\frac{\hbar v_R}{2}
\int\frac{\,dx}{2\pi} \left(\partial_x{\mathcal B}\right)^T 
{\bf 1}_2 \otimes \left( \begin{array}{cc}
1+g_4 & g_2  \\
g_2 & g+g_4 
\end{array} \right)
\left(\partial_x{\mathcal B}\right).
\end{equation}
Note that there is no coupling between the original and fictitious fields, which guarantees that the dynamics of the original fields remains unchanged. 
Using the fact of the chiral symmetry between the original and fictitious
branches we form the two couples of dual fields out of $\phi_{R,L}$ and $\phi'_{L,R}$
which read
\begin{eqnarray}
\Phi_{1}=\frac{1}{\sqrt{2}}\left(\phi'_{L}+\phi_R\right), && 
\Theta_{1}=\frac{1}{\sqrt{2}}\left(\phi'_{L}-\phi_R\right); \label{dual_1}\\
\Phi_{2}=\frac{1}{\sqrt{2}}\left(\phi_{L}+\phi'_R\right), && 
\Theta_{2}=\frac{1}{\sqrt{2}}\left(\phi_{L}-\phi'_R\right). \label{dual_2}
\end{eqnarray}
Substituting (\ref{dual_1}), (\ref{dual_2}) into the Hamiltonian (\ref{main}) we obtain 
\begin{equation}
\label{main2}
H=\frac{\hbar v_R}{2}
\int\frac{dx}{2\pi}\left[(\partial_x \Phi^T) M_\Phi (\partial_x\Phi)
+(\partial_x\Theta^T) M_\Theta (\partial_x\Theta)\right],
\end{equation}
where
\begin{equation}
M_{\Phi,\Theta}=\left( \begin{array}{cc}
1+g_4 & \pm g_2  \\
\pm g_2 & g+g_4
\end{array} \right),
\end{equation}
the upper and  lower signs corresponding to $M_{\Phi}$ and $M_{\Theta}$ respectively, 
and  $\Phi=(\Phi_1,\Phi_2)^T$, $\Theta=(\Theta_1,\Theta_2)^T$.

Hamiltonian (\ref{main}) can be brought to the canonical form applying a composition of unitary rotations and rescalings that preserve conformal invariance. We first diagonalize the matrix $M_{\Phi}=P_{\Phi}\Lambda_{\Phi}P_{\Phi}^{-1}$ by a unitary rotation of the fields with a matrix $P_{\Phi}$. Then we rescale the fields while preserving duality relations  $\left(P_{\Phi}\Phi\right)_i\rightarrow \left(P_{\Phi}\Phi\right)_i/\sqrt{\lambda^{\Phi}_i}$,
$\left(P_{\Phi}\Theta\right)_i\rightarrow \sqrt{\lambda^{\Phi}_i}\left(P_{\Phi}\Theta\right)_i$, so that after the rescaling the fields $\Phi_i$ are coupled by the unity matrix, and the Hamiltonian acquires the form 
\begin{equation}
H=\frac{\hbar v_R}{2}\int\frac{dx}{2\pi}\left[(\partial_x \Phi^T){\bf 1}(\partial_x\Phi) 
+(\partial_x\Theta^T) \tilde{M}_\Theta  (\partial_x\Theta)\right],  
\label{Htransform1}
\end{equation}
where $\tilde{M}_\Theta =\sqrt{\Lambda_{\Phi}}P^{-1}_{\Phi}M_{\Theta}P_{\Phi}\sqrt{\Lambda_{\Phi}}$. 

Further we diagonalize the quadratic form with the fields $\Theta$ by a unitary rotation with a matrix $P_{\Theta}$, i. e.
$P^{-1}_{\Theta}\tilde{M}_\Theta P_{\Theta}=\Lambda_{\Theta}$.
The unity matrix that couples the fields $\Phi$ remains unaffected by this transformation. Finally, we repeat the rescaling of the fields while preserving duality and bring the Hamiltonian to the canonical form 
\begin{equation}
\label{main3}
H=\frac{\hbar u_{1}}{2}
\int\frac{dx}{2\pi}\left(\partial_x\eta_1^2 +\partial_x\xi^2_1\right)+
\frac{\hbar u_{2}}{2}
\int\frac{dx}{2\pi}\left(\partial_x\eta_2^2 +\partial_x\xi^2_2\right),
\end{equation}
where the new dual fields $\eta=(\eta_1, \eta_2)^T$, $\xi=(\xi_1,\xi_2)^T$
are related to the original ones by transformations
$\Phi=S_\Phi \eta$, $\Theta=S_\Theta \xi$
with 
\begin{equation}
S_{\Phi}= P_{\Phi} \frac{1}{\sqrt{\Lambda_{\Phi}}} P_{\Theta} 
\sqrt[4]{\Lambda_{\Theta}}, \quad
S_{\Theta}= P_{\Phi}\sqrt{\Lambda_{\Phi}}
P_{\Theta}\frac{1}{\sqrt[4]{\Lambda_{\Theta}}}.
\label{S} 
\end{equation}
The new velocities $u_i$ are determined by $u_i=v_R\sqrt{\lambda^{\Theta}_i}$
and can be written explicitely in the form
\begin{eqnarray}
\label{u12} &&
u_{1,2}=
v_R\left(\frac{1+g^2}{2}+g_4\left(1+g+g_4\right)-g_2^2\right.\pm \\
\nonumber && 
\left. (1-g)\left\{\frac{\left[(1+g)/2+g_4\right]^2 (1-g)^2}{(1-g)^2+4g_2^2} +
\right.\right.\\
&& 
\left.\left.\left[g+g_4(1+g+g_4)-g_2^2\right]
\left[1-\frac{(1-g)^2}{(1-g)^2+4g_2^2}\right]\right\}^\frac{1}{2}
\right)^\frac{1}{2}.
\nonumber
\end{eqnarray}

{\em Tunneling density of states and tunneling current.} 
The tunneling current can be represented as a sum of partial currents between the 2DEG and the right- and left-moving branches of TLL 
\begin{equation}
I_\mathrm{pc}=I_\mathrm{pcR}+I_\mathrm{pcL} 
\label{Itotal}
\end{equation}
with the partial currents  given by  \cite{mahan} 
\begin{eqnarray}
\nonumber && 
I_{\mathrm{pc},R(L)}=\frac{4\pi e}{\hbar}\mid t \mid^2
\int d\varepsilon
\left[n_F\left(\varepsilon-eV_\mathrm{pc}\pm {eV_{\rm sd}}/{2}\right) \right.  \\
&& \left. 
-n_F(\varepsilon)\right]\nu_\mathrm{2DEG}
\nu_{R(L)}\left(\varepsilon\right),
\label{IpcR/L}
\end{eqnarray}
where $t$ is the tunneling matrix element describing
the point contact, and $n_F$ is the Fermi distribution.
The electron density of states in a 2DEG $\nu_\mathrm{2DEG}$ is just an energy-independent constant. 
The tunneling density of states $\nu_{R(L)}(\epsilon)$ is calculated from the one particle Green's function of an electron in the TLL. Using the bosonization identity $\psi_{R(L)}\sim e^{i\phi_{R(L)}}$, Eqs. (\ref{dual_1}), (\ref{dual_2}), and the relations between the initial and canonical bosonic fields given by Eqs. (\ref{S}),
we obtain the imaginary time one particle Green's function in the form  
\begin{eqnarray}
\nonumber &&
\mathcal{G}_{R(L)}(x,\tau; x, 0)=- \langle \mathcal{T}\psi_{R(L)}(x,\tau)
\psi^+_{R(L)}(x, 0)\rangle  \\
\nonumber && 
= \left\langle \mathcal{T} e^{\frac{i}{\sqrt{2}}\left(\Phi_{1(2)} \mp 
\Theta_{1(2)}\right)_{x,\tau}} e^{-\frac{i}{\sqrt{2}}\left(\Phi_{1(2)}\mp  
\Theta_{1(2)}\right)_{x,0}}\right\rangle \\
\nonumber && 
=\left[(a/(u_1\vert\tau\vert)\right]^{b_{11(21)}}   \left[a/(u_2\vert\tau\vert)\right]^{b_{12(22)}}, 
\label{G_RL}
\end{eqnarray} 
where $b_{ij}=\frac{1}{2}\left[(S_\Phi^{ij})^{2}+(S_\Theta^{ij})^{2}\right]$. 
Performing analytical continuation to real time and Fourier transformation 
\cite{mahan},
we finally obtain the chiral tunneling density of states in the form
\begin{eqnarray}
\label{DoS_int_RL}
&& 
\nu_{R(L)}(\omega)=\frac{1}{2\pi^2\,\hbar}
\frac{\left(a\vert\omega\vert\right)^{b_{11(21)}+b_{12(22)}-1}}
{u_1^{b_{11(21)}} u_2^{b_{12(22)}}} \times \\
\nonumber &&
\sin\left[\pi\left(b_{11(21)}+b_{12(22)}\right)\right]
\Gamma\left(1-b_{11(21)}-b_{12(22)}\right),
\end{eqnarray}
where $a$ is a short-length cutoff, and $\omega=\varepsilon/\hbar$. 
\begin{figure}
\includegraphics[width=\columnwidth]{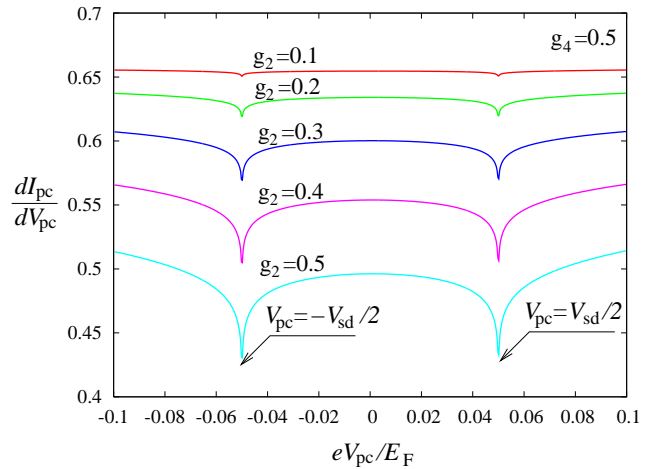}
\caption{\label{fig2} (Color online) 
Differential conductance $\partial I_\mathrm{pc}/\partial V_\mathrm{pc}$
in units of $8\pi e^2 \left\vert t\right\vert^2\nu_\mathrm{2DEG}\nu_\mathrm{1D}/\pi\hbar$
($\nu_\mathrm{1D}$ being the density of states at the Fermi level
of the unbiased noninteracting 1D channel)
as a function of voltage at the point contact for different
values of $g_2$ at a given $g_4$.
The bias voltage $V_\mathrm{sd}$ is taken equal to $0.1E_F$, and
the other parameters are relevant for typical GaAs-based electron gases:
$m^*=0.067m_e$, $a=5.65$\AA{}, $E_F=30$meV.
The conductance dependencies 
$\partial I_\mathrm{pc}/\partial V_\mathrm{sd}(V_\mathrm{pc})$
exhibit singularities at $V_\mathrm{pc}=\pm V_\mathrm{sd}/2$, 
in accordance with Eq.~(\ref{dIdV}).}
\end{figure}

The nonequilibrium tunneling density of states can be best seen in the measurements of the differential conductances $\partial I_{\mathrm {pc}}/\partial V_{\mathrm {pc,sd}}$ at small voltages $V_{\mathrm {pc}}$ on the point contact. The expression for the differential conductance can be obtained straightforwardly from Eqs. (\ref{Itotal}) and (\ref{IpcR/L}). 
The result at zero temperature can be written in the form 
\begin{eqnarray}
\nonumber && 
\frac{\partial I_{\mathrm {pc}}}{\partial V_{\mathrm {pc,sd}}} =
\zeta\frac{4\pi e^2\left\vert t \right\vert^2}{\hbar} \nu_\mathrm{2DEG}  \times \\
&&  
\left[\nu_L\left(\frac{eV_{\mathrm{sd}}}{2}+eV_{\mathrm {pc}}\right)\pm\nu_R\left(eV_{\mathrm{pc}}- \frac{eV_{\mathrm{sd}}}{2}\right)\right],  
\label{dIdV}
\end{eqnarray}
where $\zeta=1/2$ for $\partial I_{\mathrm {pc}}/\partial V_{\mathrm {sd}}$ and 
$\zeta=1$ for $\partial I_{\mathrm {pc}}/\partial V_{\mathrm {pc}}$.
In  general, the dependence of the differential conductance on $V_{\mathrm{sd}}$ and on $V_{\mathrm{pc}}$ is smooth and is determined not only by the power-law singularity in (\ref{DoS_int_RL}) but also by the dependence of $g$ on $V_{\mathrm{sd}}$. The latter makes the powers $b_{ij}$ and the velocities $u_{1,2}$ dependent on the voltage,
in accordance with Eq.~(\ref{u12}). The power-law singularities of nonequilibrium chiral densities of states can still be seen in the differential conductance at $V_{\mathrm{pc}}=\pm V_{\mathrm{sd}}/2$, as it follows from (\ref{dIdV}). At these voltages, the Fermi level in the Fermi liquid reservoir coincides with the quasi Fermi energy for the left- or the right-moving fermions in TLL, and the tunneling density of states in the corresponding channel is suppressed.
These singularities are illustrated in Fig.~\ref{fig2}
for $\partial I_\mathrm{pc}/\partial V_\mathrm{pc}$
at different values of $g_2$ which characterizes the screening
of electron-electron interactions. 
At strong screening, when $V(q=0)\simeq V(q=k_F^R-k_F^L)$ and hence 
$g_2\ll g_4$, the differential
conductance exhibits a sharp dip 
close to $V_\mathrm{pc}=\pm V_\mathrm{sd}/2$.
At weaker screening ($g_2$ closer to $g_4$) the conductance is getting smaller, and
the dip  is essentially broadened.

Furthermore, since $b_{11}=b_{22}$ and  $b_{21}=b_{12}$ for any parameters $g$ and $g_{2,4}$ the energy dependencies of  $\nu_R$ and $\nu_L$  are the same. There is however a difference in the prefactors that depend on the powers of plasmon velocities $u_1$ and $u_2$. 
This asymmetry  is maximal  at $b_{12(21)} = 0$ (i. e. when $g_2=0$). In that case the only effect of interactions consists of the renormalization of the plasmon velocities, while the singularity in the tunneling density of states disappears.   
At larger $g_2$ (i. e. $b_{12(21)}\neq 0$) the electrons with opposite chiralities
interact with each other that leads to the
alignment of the chiral tunneling densities of states $\nu_R$ and $\nu_L$, as one can
see from Eq.~(\ref{DoS_int_RL}).
\begin{figure}
\includegraphics[width=\columnwidth]{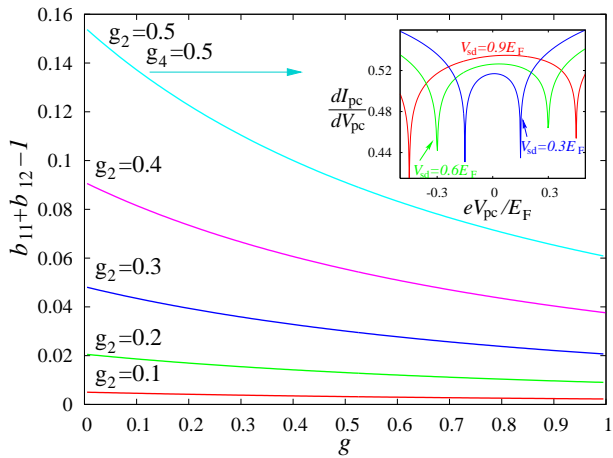}
\caption{\label{figbij} (Color online)
The dependence of the tunneling exponent $\alpha=b_{11}+b_{12}-1=b_{21}+b_{22}-1$ on the 
asymmetry of the Fermi velocities $v_R$ and $v_L$ expressed through the parameter $g=v_L/v_R$.
The unbiased TLL corresponds to $g=1$. The asymmetry increases ($g$ diminishes)  increasing the bias voltage $V_{\rm sd}$. The suppression of the tunneling density of states grows with bias voltage. Inset: Differential conductance 
$\partial I_\mathrm{pc}/\partial V_\mathrm{pc}$
as a function of voltage at the point contact for different
values of $V_\mathrm{sd}$ at a given $g_2=g_4=0.5$.
The other parameters are taken the same as for Fig.~\ref{fig2}.}
\end{figure}
The dependence of the tunneling exponent $\alpha=b_{11}+b_{12}-1=b_{21}+b_{22}-1$  on the asymmetry of the Fermi velocities is shown in Fig. \ref{figbij}. The power $\alpha$ increases  with the grows of asymmetry (smaller $g$)  from its equilibrium value at $g=1$. Therefore, the singularity in the tunneling density of states becomes stronger with increase of the bias voltage $V_{\rm sd}$.
It is interesting that the finite size of the TLL also leads to increasing of
the exponent of a power in the tunneling density of states \cite{eggert}.
Note, however, that it is much easier to change the bias voltage than the length of the wire.

Another interesting feature of the system considered
is the finite current through the point contact
even at vanishing $V_\mathrm{pc}$. 
This is again due to the chiral asymmetry of the plasmon spectrum
subject to the bias voltage $V_\mathrm{sd}$. 
The existence of a finite current follows directly from Eq.~(\ref{dIdV}), but
in oder to ease the understanding of that fact, we rewrite
$\partial I_\mathrm{pc}/\partial V_\mathrm{sd}$
in the limiting case of $V_\mathrm{pc}=0$ and strong screening when
$g_2=0$. The latter leads to
$b_{12}=b_{21}=0$, $b_{11}=b_{22}=1$,
$u_1=v_R(1+g_4)$, $u_2=v_L+v_R g_4$,
and the differential conductance assumes the form
\begin{equation}
\label{simplified}
\frac{\partial I_{\mathrm {pc}}}{\partial V_{\mathrm {sd}}} =
\frac{e^2}{\hbar^2}
\frac{\left\vert t \right\vert^2\nu_\mathrm{2DEG} (v_R-v_L)}
{(v_L+v_R g_4)(v_R+v_R g_4)}.
\end{equation}
From Eq.~(\ref{simplified}) it is clear
that, on one hand, the tunneling current
into the TLL can be suppressed due to the electron-electron interactions.
On the other hand, one can facilitate the tunneling
applying the bias voltage $V_\mathrm{sd}$ to the TLL.
We emphasize that one does not need to change $V_\mathrm{pc}$. 
Since high voltages at the point contact are not always
possible in the linear response measurements, 
the biasing of the TLL might be a powerfull tool to study its transport
properties.

{\em Conclusions.}
In conclusion, we showed that in the experiment on the tunneling into a strongly biased TLL through a point contact, the power law singularities in the tunneling densities of states can be seen separately for the right- and left-movers. We obtained analytical expressions for the chiral tunneling densities of states that turn out to be different from the equilibrium case.
The predicted behavior can be observed in experiments with GaAs quantum wires of nominal width $\sim 14 \  {\rm nm}$, where the application of a strong bias voltage does not cause the population of the next one-dimensional subband. Fabrication of such wires lies within the range of current experiments \cite{experiment-semi,experiment-wire}.
We also developed a method of diagonalization of TLL Hamiltonian with different Fermi velocities for the right- and left-movers. The method can be usefull for a number of problems which involve chiral asymmetry of the density of states
such as a TLL wire in an external magnetic field.

\begin{acknowledgments} 
M. T. is grateful to Milena Grifoni for fruitful discussions and useful advises.
A. C. is grateful to L. Glazman,  A. Kamenev, M. Khodas, D. Novikov,
and Y. Adamov for illuminating discussions. A. C. appreciates and enjoyed the hospitality  of the William I. Fine Theoretical Physics Institute, University of Minnesota, where a part of this work has been performed. 
The authors acknowledge financial support from DFG through Graduiertenkolleg ``Physik nanostrukturierter Festk\"orper''  (M.T.) and Sonderforschungsbereich 508 (A. C.). 

\end{acknowledgments}


\begin{thebibliography}{99}

\bibitem{precursors}{C.~L. Kane and M.~P.~A. Fisher,
Phys. Rev. B \textbf{46}, 7268 and 15233 (1992),
Enss et al, {\em ibid.} \textbf{71}, 155401 (2005),
I. Ussishkin and L.~I. Glazman, Phys. Rev. Lett. \textbf{93}, 196403 (2004),
S. Eggert {\em ibid.} \textbf{84}, 4413 (2000),
H.~G. S. Huegle, R. Egger and H. Grabert, Sol. State Comm.
\textbf{117}, 93 (2000).}

\bibitem{experiment-carbon}{M. Bockrath et al, Nature \textbf{397}, 598 (1999),
H.~W.~Ch. Postma et al, Science \textbf{293}, 76 (2001).}

\bibitem{experiment-semi}{O. M. Auslaender et al, Science \textbf{295}, 825 (2002).}

\bibitem{classics}{S. Tomonaga, Prog. Theor. Phys. \textbf{5}, 544 (1950),
J.~M. Luttinger, J. Math. Phys. \textbf{4}, 1154 (1963),
R. Heidenreich et al, Phys. Lett. A \textbf{54}, 119 (1975),
F.~D.~M. Haldane, J. Phys. C \textbf{14}, 2585 (1981).}

\bibitem{vonDelft}{J. von Delft and H. Schoeller,
Ann. der Physik \textbf{7}, 225 (1998).}

\bibitem{curvature}{E. Bettelheim, A.~G. Abanov, and P. Wiegmann, 
Phys. Rev. Lett. \textbf{97}, 246402 (2006),
M. Pustilnik et al, {\em ibid.} \textbf{96}, 196405 (2006).}

\bibitem{gogolin}{A. Gogolin, A. Nersesyan,
and A. Tsvelik,
  \emph{\bibinfo{title}{Bosonization in Strongly Correlated Systems}}
  (\bibinfo{publisher}{Cambridge University Press}, \bibinfo{year}{1998}).}

\bibitem{mahan}{G. Mahan,
  \emph{\bibinfo{title}{Many-Particle Physics}} (\bibinfo{publisher}{Plenum
  Press, New York}, \bibinfo{year}{1981}).}

\bibitem{eggert}{S. Eggert, H. Johannesson,
and A. Mattsson, Phys. Rev. Lett. \textbf{76},
1505 (1996).}

\bibitem{experiment-wire}{K. Wagner et al.
Phys. Rev. Lett. \textbf{97}, 056803 (2006).}


\end{thebibliography}
\end{document}